\documentclass[aps,prd,twocolumn,superscriptaddress,showpacs,letter]{revtex4-1}            
\usepackage[utf8]{inputenc}
\usepackage{graphicx}
\usepackage{dcolumn}
\usepackage{bm}
\usepackage{cancel}
\usepackage{color}             
\usepackage{epsfig}
\usepackage{amssymb}
\usepackage{amsmath}
\usepackage{multirow}
\usepackage{hyperref}
\usepackage{cleveref}

\begin{document}

\title{Unitarity effects in high-energy elastic scattering}
\thanks{Presented at ``New Trends in High-Energy and Low-$x$ Physics'', Sfantu Gheorghe, Romania, September 1-5, 2024.}

\author{M.~Maneyro}
\email{maneyro@fing.edu.uy}
\affiliation{Instituto de F\'{\i}sica, Facultad de Ingenier\'{\i}a, Universidad de la Rep\'ublica, \\
  J.H. y Reissig 565, 11000 Montevideo, Uruguay}
\affiliation{Department of Physics, University of Liverpool, \\ L69 7ZE, Liverpool, United Kingdom}
\author{E.~G.~S.~Luna}
\email{luna@if.ufrgs.br}
\affiliation{Instituto de F\'isica, Universidade Federal do Rio Grande do Sul, Caixa Postal 15051, 91501-970, Porto Alegre, Rio Grande do Sul, Brazil}
\author{M.~Pel\'aez}
\email{mpelaez@fing.edu.uy}
\affiliation{Instituto de F\'{\i}sica, Facultad de Ingenier\'{\i}a, Universidad de la Rep\'ublica, \\
J.H. y Reissig 565, 11000 Montevideo, Uruguay}
 

\begin{abstract}

We investigate the high-energy behavior of the elastic scattering amplitude using the eikonal and $U$-matrix unitarization schemes. This work extends the analysis in \cite{mlp001} by exploring the sensitivity of the Pomeron and Odderon parameters to the inclusion of differential cross-section data over an extended range of $|t|$.

\end{abstract}


\maketitle

\section{Introduction}

In a recent study \cite{mlp001}, we studied the high-energy behavior of the elastic scattering amplitude using two distinct unitarization schemes: the eikonal and the  $U$-matrix. In that analysis, which examined the properties of both the  Pomeron and Odderon, we found that the Odderon phase factor $\xi_{\Bbb  O} = -1$ is favored in both schemes, indicating an Odderon with a phase opposite to other odd-signature components of the scattering amplitude. This result was obtained after a global analysis of two data ensembles (one comprising data from the ATLAS Collaboration and the other from the TOTEM Collaboration) covering total cross-sections, the $\rho$  parameter, and differential cross-section data for $pp$ scattering within the range $|t| \leq 0.1$ GeV$^{2}$. This work uses the same formalism and statistical methods as in \cite{mlp001} but extends the analysis to ensembles that include $d\sigma/dt$ data over the broader range $|t| \leq 0.2$ GeV$^{2}$. Our goal is to assess the sensitivity of the Pomeron and Odderon parameters to the inclusion of $d\sigma/dt$ data with this wider momentum interval.

\section{Formalism}

The construction of unitarized scattering amplitudes involves some formal steps. First, we define a Born-level amplitude, ${\cal F}(s,t)$, with its even- and odd-crossing components expressed as
\begin{eqnarray}
{\cal F}^{\pm}(s,t) = \frac{1}{2} \left[ {\cal F}^{pp}(s,t) \pm  {\cal F}^{\bar{p}p}(s,t) \right] .
\label{eqn01} 
\end{eqnarray}
The corresponding Born amplitudes for even- and odd-crossing in impact parameter space, $b$, are given by
\begin{eqnarray}
\chi^{\pm}(s,b) = \frac{1}{s}\int_{0}^{\infty} q \, dq \, J_{0}(bq) {\cal F}^{\pm}(s,-q^{2}) ,
\label{eqn02}
\end{eqnarray}
where $-q^{2}=t$. Next, we construct a new scattering amplitude, $H^{pp}_{\bar{p}p}(s,b)$, in terms of the functions $\chi^{pp}_{\bar{p}p}(s,b)$. Finally, the amplitude in momentum space, which is used for computing observables, is obtained through the inverse Fourier-Bessel transform of $H^{pp}_{\bar{p}p}(s,b)$:
\begin{eqnarray}
{\cal A}^{pp}_{\bar{p}p}(s,t) = s \int_{0}^{\infty}b\, db\, J_{0}(bq)\, H^{pp}_{\bar{p}p}(s,b) .
\label{scatter007}
\end{eqnarray}
In the eikonal scheme (Es), the relation
\begin{eqnarray}
H(s,b) = i \left[ 1 - e^{i\chi (s,b)}  \right] 
\label{eikrelation01}  
\end{eqnarray}
holds, leading to
\begin{eqnarray}
{\cal A}_{[Es]}(s,t) = is \int^{\infty}_{0} b\, db\, J_{0}(bq) \left[ 1 - e^{i\chi (s,b)} \right] .
\end{eqnarray}
In contrast, in the $U$-matrix scheme (Us), the relation
\begin{eqnarray}
H(s,b) = \frac{\chi(s,b)}{1-i\chi(s,b)/2} 
\label{unitarz01}
\end{eqnarray}
implies
\begin{eqnarray}
{\cal A}_{[Us]}(s,t) = i s \int^{\infty}_{0} b\, db\, J_{0}(bq) \left[ \frac{2 \chi(s,b)}{\chi(s,b) + 2i} \right] .
\end{eqnarray}

The input amplitudes ${\cal F}_{i}(s,t)$ are associated with Reggeon exchange amplitudes. The Born amplitude for each single exchange is
\begin{eqnarray}
{\cal F}_{i}(s,t) = \beta_{i}^{2}(t)\eta_{i}(t)\left( \frac{s}{s_{0}} \right)^{\alpha_{i}(t)} ,
\label{equation05}
\end{eqnarray}
$i=-, +, \Bbb O, \Bbb P$, where $\beta_{i}(t)$ is the elastic proton-Reggeon vertex, $\eta_{i}(t)$ is the signature factor, $\alpha_{i}(t)$ is the Regge trajectory, and  $s_{0} \equiv 1$ GeV$^{2}$ is an energy scale. Here ${\cal F}_{-}(s,t)$ represents the exchange of Reggeons having $C=-1$ parity ($\omega$ and $\rho$), ${\cal F}_{+}(s,t)$ the exchange of
Reggeons having $C=+1$ parity ($a_{2}$ and $f_{2}$), ${\cal F}_{\Bbb O}(s,t)$ represents the $C=-1$ Odderon exchange, while ${\cal F}_{\Bbb P}(s,t)$ represents the $C=+1$ Pomeron exchange.

For Regge trajectories with odd-signature, the signature factor is given by $\eta_{i}(t) = -ie^{-i\frac{\pi}{2}\alpha_{i}(t)}$, while for even-signature trajectories, it is $\eta_{i}(t) = -e^{-i\frac{\pi}{2}\alpha_{i}(t)}$. Reggeons with positive charge-conjugation are assumed to have an exponential form for the proton-Reggeon vertex,
\begin{eqnarray}
\beta_{+}(t)=\beta_{+}(0)\exp (r_{+}t/2 ) ,
\end{eqnarray}
and lie on a linear trajectory of the form
\begin{eqnarray}
\alpha_{+}(t) = 1 - \eta_{+} + \alpha^{\prime}_{+} t.
\end{eqnarray}
Similarly, Reggeons with negative charge-conjugation are characterized by the parameters
$\beta_{-}(0)$, $r_{-}$, $\eta_{-}$, and $\alpha^{\prime}_{-}$.

The input amplitude for the Odderon contribution is expressed as
\begin{eqnarray}
{\cal F}_{\Bbb O}(s,t) =  \beta_{\Bbb O}^{2}(t)\, \eta_{\Bbb O}(t) \left( \frac{s}{s_{0}} \right)^{\alpha_{\Bbb O}(t)} ,
\label{odd1}
\end{eqnarray}
where $\eta_{\Bbb O}(t) = -ie^{-i\frac{\pi}{2}\alpha_{\Bbb O}(t)}$ and $\alpha_{\Bbb O}(t) = 1$. We adopt a power-like form for the proton-Odderon vertex, given by
\begin{eqnarray}
\beta_{\Bbb O}(t)=\frac{\beta_{\Bbb O}(0)}{(1-t/m_{\rho}^{2})(1-t/a_{\Bbb O})} ,
\label{vertex03}
\end{eqnarray}
where $a_{\Bbb O} = 2 a_{\Bbb P}$.

In the case of Pomeron exchange, the trajectory is given by
\cite{anselm001,kmr001,kmr002,kmr003,broilo001,lrk001},
\begin{eqnarray}
\alpha_{\Bbb P}(t) = \alpha_{\Bbb P}(0) + \alpha^{\prime}_{\Bbb P} t + \frac{m_{\pi}^{2}}{32\pi^{3}}\,
h(\tau) ,
\label{pomnlin}
\end{eqnarray}
where $\eta_{\Bbb P}(t) = -e^{-i\frac{\pi}{2}\alpha_{\Bbb P}(t)}$, $\alpha_{\Bbb P}(0) = 1 + \epsilon$, and
\begin{eqnarray}
h (\tau) &=& -\frac{4}{\tau}\, F_{\pi}^{2}(t) \left[  2\tau - (1+\tau)^{3/2} \ln \left( \frac{\sqrt{1+\tau}+1}{\sqrt{1+\tau}-1} \right)
\right. \nonumber \\
& & + \left. \ln \left( \frac{m^{2}}{m_{\pi}^{2}} \right) \right]  ,
\label{nonlinear01}
\end{eqnarray}
with $\epsilon > 0$, $\tau = 4m_{\pi}^{2}/|t|$, $m_{\pi}=139.6$ MeV, and $m=1$ GeV. In the equation above, $F_{\pi}(t)$ is the form
factor of the pion-Pomeron vertex, given by $F_{\pi}(t)=\beta_{\pi}/(1-t/a_{1})$; the coefficient $\beta_{\pi}$ specifies the value of the pion-Pomeron coupling, where we take the relation $\beta_{\pi}/\beta_{I\!\!P}(0)=2/3$. The proton-Pomeron vertex also takes a power-like form,
\begin{eqnarray}
\beta_{\Bbb P}(t)=\frac{\beta_{\Bbb P}(0)}{(1-t/a_{1})(1-t/a_{\Bbb P})} ,
\label{vertex02}
\end{eqnarray}
where the free parameter $a_{1}=m_{\rho}^{2} = (0.776\, \textnormal{GeV})^{2}$ in (\ref{vertex02}) is the same as the one in the expression for $F_{\pi}(t)$.

The corresponding amplitudes $\chi_{i}(s,b)$ are obtained by applying a Fourier-Bessel transform applied to each ${\cal F}_{i}(s,t)$:
\begin{eqnarray}
\chi_{i}(s,b) = \frac{1}{s}\int \frac{d^{2}q}{2\pi}\, e^{i{\bf q}\cdot {\bf b}} \, {\cal F}_{i}(s,t) .
\label{gtgt01}
\end{eqnarray}
The physical amplitudes in $b$-space are obtained by summing the Fourier-Bessel transforms of all possible exchanges, expressed as
\begin{eqnarray}
\chi^{pp}_{\bar{p}p} (s,b) =  \chi_{\Bbb P}(s,b)  + \chi_{+}(s,b)  \pm \chi_{-}(s,b) \pm \xi_{\Bbb O}\chi_{\Bbb O}(s,b) , \nonumber \\
\label{eikonal009}
\end{eqnarray}
where $\chi_{\Bbb O}(s,b)$ represents the Odderon's phase factor, associated with the positivity property. We consider an Odderon with a phase factor $\xi_{\Bbb O}= -1$ \cite{mlp001}.

The total cross section, elastic differential cross section, and the $\rho$ parameter are expressed in terms of the physical
amplitude ${\cal A}^{pp}_{\bar{p}p}(s,t)$,
\begin{eqnarray}
\sigma^{pp, \bar{p}p}_{tot}(s)=\frac{4\pi}{s}\, \textnormal{Im}\, {\cal A}^{pp}_{\bar{p}p}(s,t=0) ,
\label{ertion001}
\end{eqnarray}
\begin{eqnarray}
\frac{d\sigma^{pp, \bar{p}p}}{dt}(s,t)=\frac{\pi}{s^{2}}\, \left| {\cal A}^{pp}_{\bar{p}p}(s,t) \right|^{2} ,
\label{ertion002}
\end{eqnarray}
\begin{eqnarray}
\rho^{pp, \bar{p}p}(s)=\frac{\textnormal{Re}\, {\cal A}^{pp}_{\bar{p}p}(s,t=0)}{\textnormal{Im}\, {\cal A}^{pp}_{\bar{p}p}(s,t=0)} .
\label{ertion003}
\end{eqnarray}
In the eikonal scheme, we replace ${\cal A}^{pp}_{\bar{p}p}(s,t)$ with ${\cal A}^{pp, \bar{p}p}_{[Es]}(s,t)$, such that
\begin{eqnarray}
{\cal A}^{pp, \bar{p}p}_{[Es]}(s,t) = is \int^{\infty}_{0} b\, db\, J_{0}(bq) \left[ 1 - e^{i\chi^{pp}_{\bar{p}p} (s,b)} \right] .
\end{eqnarray}
In the $U$-matrix scheme, we similarly replace ${\cal A}^{pp}_{\bar{p}p}(s,t)$ with ${\cal A}^{pp, \bar{p}p}_{[Us]}(s,t)$, leading to
\begin{eqnarray}
{\cal A}^{pp, \bar{p}p}_{[Us]}(s,t) = i s \int^{\infty}_{0} b\, db\, J_{0}(bq) \left[ \frac{2 \chi^{pp}_{\bar{p}p}(s,b)}{\chi^{pp}_{\bar{p}p}(s,b) + 2i} \right] .
\end{eqnarray}

\section{Results and Conclusions}

\begin{table*}
\centering
\caption{The Pomeron, Odderon and secondary Reggeons parameters values obtained in global fits to Ensembles A and T after the Eikonal and $U$-matrix unitarizations.}
\begin{ruledtabular}
\begin{tabular}{ccccc}
 & \multicolumn{2}{c}{Eikonal unitarization} & \multicolumn{2}{c}{$U$-matrix unitarization} \\
\cline{2-3} \cline{4-5} 
 & Ensemble A & Ensemble T & Ensemble A & Ensemble T  \\
\hline
$\epsilon$ & 0.1172$\pm$0.0053 & 0.1232$\pm$0.0025 & 0.0968$\pm$0.0015 & 0.1146$\pm$0.0038 \\
$\alpha^{\prime}_{I\!\!P}$ (GeV$^{-2}$) & 0.0109$\pm$0.0050 & 0.0109$\pm$0.0037 & 0.2624$\pm$0.0042 & 0.092$\pm$0.011   \\
$\beta_{\Bbb P}(0)$ & 1.918$\pm$0.082 & 1.895$\pm$0.041 & 2.162$\pm$0.025 & 1.923$\pm$0.067  \\
$a_{\Bbb P}$ (GeV$^{-2}$) & 0.47$\pm$0.11 & 0.53$\pm$0.24 & 40$\pm$24 & 0.585$\pm$0.051  \\
$\beta_{\Bbb O}(0)$ & 0.35$\pm$0.21 & 0.27$\pm$0.18 & 0.31$\pm$0.24 & 0.28$\pm$0.17  \\
$\eta_{+}$ & 0.324$\pm$0.054 & 0.324$\pm$0.029 & 0.374$\pm$0.017 & 0.317$\pm$0.051  \\
$\beta_{+}(0)$ & 4.23$\pm$0.47 & 4.23$\pm$0.25 & 4.51$\pm$0.12 & 4.15$\pm$0.43  \\
$\eta_{-}$ & 0.48$\pm$0.11 & 0.48$\pm$0.14 & 0.476$\pm$0.090 & 0.499$\pm$0.078  \\
$\beta_{-}(0)$ & 3.09$\pm$0.63 & 3.09$\pm$0.53 & 3.08$\pm$0.73 & 3.16$\pm$0.48  \\
\hline
$\nu$ & 264 & 496 & 264 & 496  \\
\hline
$\chi^{2}/\nu$ & 0.75 & 1.40 & 0.88  & 0.60  \\
\end{tabular}
\end{ruledtabular}
\label{tab001}
\end{table*}

\begin{figure*}\label{fig003}
\begin{center}
\includegraphics[height=.80\textheight]{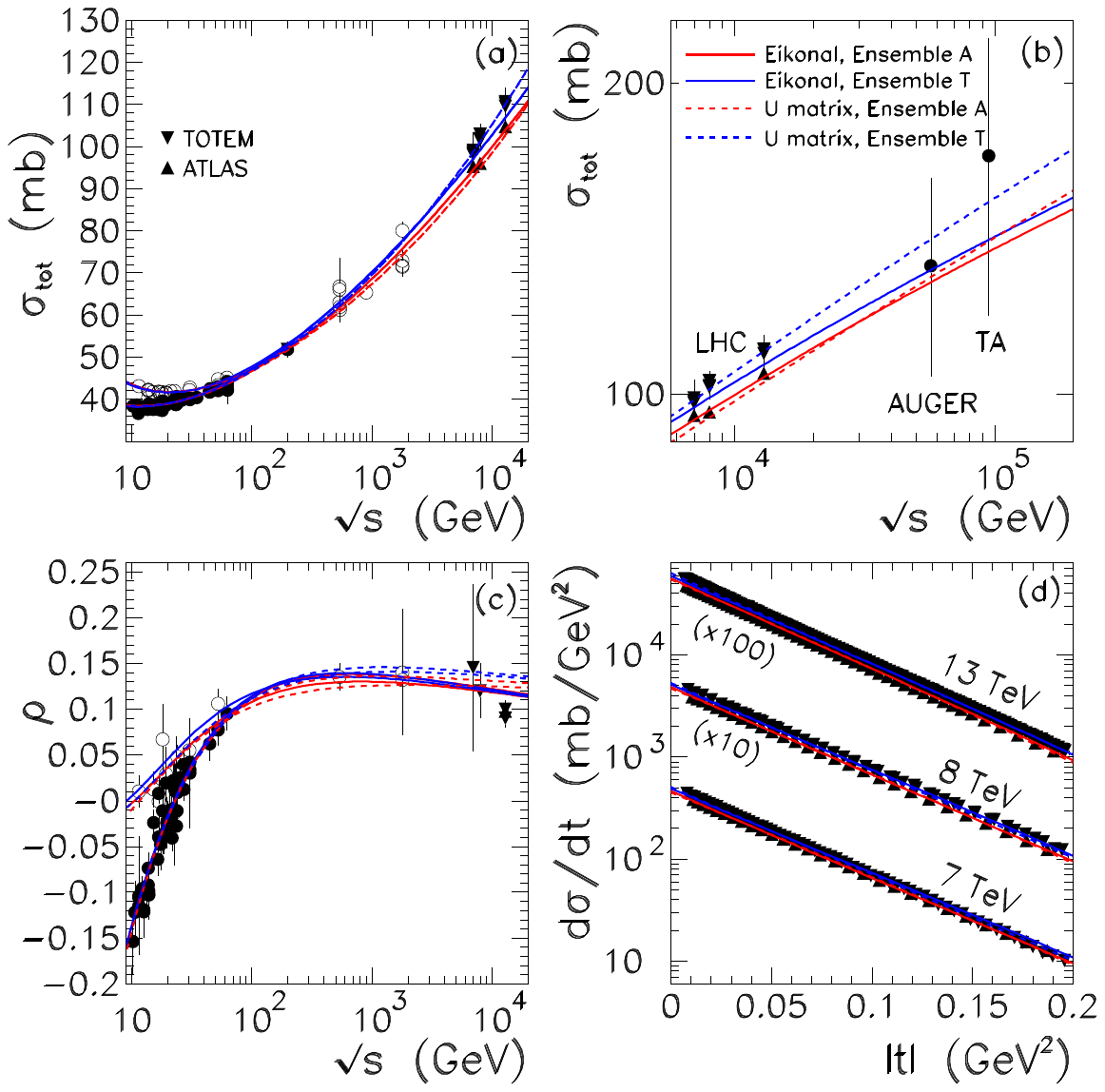}
\caption{Total cross section, $\rho$ parameter, and differential cross section for $pp$ ($\bullet$, $\blacktriangle$, $\blacktriangledown$) and $\bar{p}p$ ($\circ$) channels.}
\end{center}
\end{figure*}

We perform global fits to the total cross section, $\sigma_{tot}^{pp,\bar{p}p}$, and to the ratio of the real to the imaginary part of the scattering amplitude, $\rho^{pp,\bar{p}p}$, above $\sqrt{s} = 10$ GeV and to $pp$ differential cross-section data while considering two distinct data ensembles, one with ATLAS measurements and the other with TOTEM measurements (there is a noteworthy tension between the TOTEM and ATLAS measurements \cite{petrov01,petrov02,bopsin01}).
Precisely, we fit to the elastic differential cross section, $d\sigma^{pp}/dt$, at LHC with $|t| \leq 0.2$ GeV$^{2}$. We use $\sigma_{tot}$ and $\rho$ data compiled by the Particle Data Group \cite{pdg001}, and $d\sigma/dt$ data at 7, 8, and 13 TeV from the TOTEM \cite{antchev001,antchev002,TOTEM001,TOTEM005,TOTEM008,TOTEM010} and ATLAS \cite{atlas001,atlas002,ATLAS01} Collaborations. For all the data, the statistical and systematic errors were added in quadrature.

The two data ensembles are defined as follows:

{\bf Ensemble T}: $\sigma_{tot}^{pp,\bar{p}p}$ data + $\rho^{pp,\bar{p}p}$ data + TOTEM data on $\frac{d\sigma}{dt}$ at 7, 8, and 13 TeV;

{\bf Ensemble A}: $\sigma_{tot}^{pp,\bar{p}p}$ data + $\rho^{pp,\bar{p}p}$ data + ATLAS data on $\frac{d\sigma}{dt}$ at 7, 8, and 13 TeV.

The Ensemble T (A) includes $\sigma_{tot}^{pp}$ and $\rho^{pp}$ data measured by the TOTEM (ATLAS) collaboration. 

The procedure of separating discrepant data collected at the same center-of-mass energy to generate two distinct data ensembles is statistically robust. It has previously been employed to investigate discrepancies in cosmic-ray data and their impact on predictions for $pp$ total cross sections at high energies \cite{luna011}. This approach has also been applied to analyze the tension between CDF and E710/E811 Tevatron data and its effects on determining extreme bounds for the Pomeron intercept \cite{luna012a,luna012b,luna012c}.

Having clearly defined our data sets, we proceed to the phenomenological analysis by performing global fits on the two distinct ensembles. We employ a $\chi^{2}$ fitting procedure, where the minimum $\chi^{2}_{min}$ is expected to follow a $\chi^{2}$ distribution with $\nu$ degrees of freedom. The global fittings for the experimental data incorporate a $\chi^{2}$ interval that, in the presence of normal errors, corresponds to the region covering 90\% of the probability within the $\chi^{2}$ hypersurface. This translates to $\chi^{2}-\chi^{2}_{min}=13.36$ and $14.68$ for models with eight and nine free parameters, respectively.
To minimize the number of free parameters, we fix the slopes of the secondary-Reggeon linear trajectories, $\alpha^{\prime}_{+}$ and $\alpha^{\prime}_{-}$, at 0.9 GeV$^{-1}$, consistent with typical values observed in Chew-Frautschi plots. Furthermore, we set the slopes related to the form factors of the secondary Reggeons to $r_{+} = r_{-} = 4.0$ GeV$^{-2}$. These parameters show minimal statistical correlation with the Odderon (and Pomeron) parameters, and their fixed values align with those obtained in prior studies \cite{goulianos001,kmr002,kmr003}.

The parameters for the Odderon, Pomeron, and secondary Reggeons, determined from global fits to Ensembles A and T within the frameworks of the eikonal and $U$-matrix unitarization schemes, are presented in Table \ref{tab001}. The results of these fits are shown in Figure 1, which provides a comparison of $pp$ and $\bar{p}p$ data for both unitarization schemes. Specifically, part (a) depicts the total cross-section, part (b) extends the range of $\sqrt{s}$ shown in part (a), part (c) highlights the $\rho$ parameter (the ratio of the real to imaginary parts of the scattering amplitude), and part (d) focuses on the differential cross-section. For comparison, part (b) also includes estimates of the $pp$ total cross-section from cosmic ray experiments, namely the AUGER result at $\sqrt{s} = 57$ TeV \cite{auger} and the Telescope Array result at $\sqrt{s} = 95$ TeV \cite{TA}.

Similar to the analysis in Ref. \cite{mlp001}, we observe that the parameters associated with secondary Reggeons are not sensitive to the choice of unitarization scheme, as their values are consistent across different schemes, accounting for the associated uncertainties.

The results obtained using $U$-matrix unitarization are consistent with the previous analysis in Ref. \cite{mlp001}. The parameter values showed minimal variation, and the $\chi^{2}/\nu$ values remained low, indicating that the $d\sigma/dt$ data in the newly extended interval $|t| \leq 0.2$ GeV$^{2}$continue to be well described by the model. Once again, Ensemble A favors a higher value of $\alpha_{I\!\!P}$ compared to Ensemble T, and the smaller value of $\epsilon$ in Ensemble A is compensated by a larger value of the coupling $\beta_{I\!\!P}(0)$.

On the other hand, the results using eikonal unitarization indicate that Ensemble A now favors a small value of $\alpha^{\prime}_{\Bbb P}$. Interestingly, this value is identical in both ensembles, with $\alpha^{\prime}_{\Bbb P}\sim 0.011$ GeV$^{-2}$. It is important to note that this behavior aligns with results from screened Regge models, which also yield very small values for $\alpha^{\prime}_{\Bbb P}$ \cite{kmr002,kmr003,kmrepj01,kmrepj02,lev01,lev02}. In the framework of Gribov's Reggeon calculus, the small value of $\alpha^{\prime}_{\Bbb P}$ suggests that the soft Pomeron can potentially be treated perturbatively \cite{gribov01,gribov02,gribov03,baker01}, opening the possibility of developing a fundamental theory of soft processes based entirely in QCD.
Similar to $U$-matrix unitarization, in eikonalization, an increase in the Pomeron intercept (as seen in Ensemble T compared to Ensemble A) is offset by a decrease in the coupling parameter $\beta_{\Bbb P}(0)$.

Thus, we find that for Ensembles A and T, the eikonal unitarization scheme is sensitive to the input data for $d\sigma/dt$ when $|t| > 0.1$ GeV$^{2}$.

\section*{Acknowledgment}

This research was partially supported by the Agencia Nacional de Investigaci\'on e Innovaci\'on under the project ANII-FCE-166479 and by the Conselho Nacional de Desenvolvimento Cient\'{\i}fico e Tecnol\'ogico under Grant No. 307189/2021-0.

\end{document}